\documentclass{kluwer}    % Specifies the document style.

\newdisplay{guess}{Conjecture}

\begin{document}                                                                                   
\begin{article}
\begin{opening}         
\title{Dynamical evolution driven by bars and interactions : Input
            from numerical simulations} 
\author{E.  \surname{Athanassoula}}  
\runningauthor{E.  Athanassoula}
\runningtitle{Dynamical evolution}
\institute{Observatoire de Marseille,\\
2 Place Le Verrier,\\
13248 Marseille cedex 04,\\
France}
%\date{April 15, 1993}

\begin{abstract}
We discuss the evolution of a disc galaxy due to the formation of a
bar and, subsequently, a peanut.  After the formation stage there is
still considerable evolution, albeit slower.  In purely stellar cases
the pattern speed of the bar decreases with time, while its amplitude
grows.  However, if a considerable gaseous component is present
in the disc, the pattern speed may increase with time, while the bar
strength may decrease.  In some cases
the gas can be brought sufficiently close to the center to create a
strong central concentration, which, in turn, may modify the
properties of the bar.  

More violent evolution can take place during interactions, so that some
disc substructures can be either formed or destroyed in a time scale
which is small compared to a Hubble time.  These include
spirals, bars, bridges, tails, rings, thick discs and bulges.  In some
cases interactions may lead to mergings.  We briefly review comparisons
of the properties of merger remnants with those of elliptical galaxies,
both for the case of pairwise mergings and the case of multiple mergings.
\end{abstract}

\end{opening}           

\section{Introduction}  

Many studies, and particularly those based on numerical simulations
including time-dependence, have shown that disc galaxies can be
considered as static, or even quasi-stationary, only in a zeroth order
approximation and in fact undergo considerable
evolution with time.  We will here discuss first the evolution linked to a
bar, namely its formation, the formation of a peanut and the
subsequent slower secular evolution, which leads to changes in the bar
pattern speed and strength (section 2).  We will then turn to
the more violent evolution which occurs during interactions and
mergings (section 
3).  In particular we will discuss the formations of tails and bridges
(section 3.1), the formation of collisional ring galaxies (section
3.2), the interaction of a barred galaxy with a small spheroidal
companion (section 3.3), and the properties of merger remnants,
including those of multiple mergers (section 3.4).

\section{Bars as engines of secular evolution}

\subsection{Bar formation}

The early seventies brought the realization that galactic discs 
are unstable to the formation of a bar, unless specific care is 
taken to stabilise them, or rather to reduce the growth rate of 
the instability (e.g. \citeauthor{A84}, \citeyear{A84} or 
\citeauthor{SW}, \citeyear{SW} for reviews).  As it grows, the bar brings
considerable changes  
to the orbital structure of the disc.  In an axisymmetric disc 
with not too high radial velocity dispersion, the orbits can be 
considered quasi-circular, with deviations that can approximated
by epicycles \cite{BT}.  After the growth of the bar most stars in the bar 
region are trapped around $x_1$ type orbits, i.e. around periodic
orbits elongated  
along the bar (see e.g. \citeauthor{CG}, \citeyear{CG}).  Such orbits 
can not extend beyond corotation (hereafter CR), thus setting a 
natural limit to the length of the bar \cite{C80}.  Furthermore, the outermost
parts of the bar, just within corotation, are mainly populated
by chaotic orbits and by orbits trapped around periodic orbits of 
type $n$:1 (i.e. having $n$ radial oscillations for one rotation 
around the galactic center).  Thus the $x_1$ orbits may fall short
before CR.  This led \citeauthor{SPA2} \shortcite{SPA2} to argue that the
bar should  
stop at the 4:1 resonance, or inner Ultra-harmonic resonance 
(hereafter iUHR).  \citeauthor{A92} \shortcite{A92} found from
hydrodynamical simulation that the gas response agrees with
observations only if the CR radius is 1.2 $\pm$ 0.2 times the length
of the bar semi-major axis, and this limit is in good agreement with whatever
observational constraints are available \cite{A92, E96}.  
In fact the two criteria could well give the same limits for 
the bar length, although it is not easy to calculate the
iUHR radius for an object as far from axisymmetry as a strongly barred
disc galaxy.

After the bar has become sufficiently strong it will in turn be
prey to another instability, the buckling instability, which will
bend the orbits and bring them well outside the equatorial plane.
As a result of this the bar will thicken and take a peanut shape.
This is clearly seen in 3D simulations of bar unstable discs (e.g. 
\citeauthor{CS}, \citeyear{CS}; \citeauthor{CDFP}, \citeyear{CDFP};
\citeauthor{RSJK}, \citeyear{RSJK}).  Again there is a considerable 
change in the orbital structure after the peanut has grown.  While the 
$x_1$ family is the backbone of thin or 2D bars, it is the whole 
$x_1$ tree, including the vertical families of periodic orbits 
bifurcating from the $x_1$ at various vertical resonances, that 
is the backbone of 3D bars and peanuts \cite{SPA1}.      
  
\subsection{The role of the halo}

The halo had initially been thought to stabilise the disc and to postpone,
if not stop, the growth of the bar (\citeauthor{A84}, \citeyear{A84};
\citeauthor{SW}, \citeyear{SW}; and references in either).  Yet 
the results in \citeauthor{AM} \shortcite{AM} and \citeauthor{Atokyo}
\shortcite{Atokyo} show strong bars in simulations in which the disc
is initially embedded in a very massive halo.  Athanassoula
\shortcite{Adunk} suggested that this is due to the response of the
halo.  Indeed, simulations with rigid haloes, whose mass 
within the disc region is a large fraction of the total mass within
that volume, show that no strong bar is formed in a Hubble time; at
the most a small oval in the 
central parts.  On the other hand, if the halo is live it responds to
the bar and takes angular momentum from it.  This excites the bar,
since the latter is a negative angular momentum `perturbation', and makes it
grow stronger.  In particular, there is a considerable fraction of
resonant particles/orbits in the halo component.
  
\subsection{Evolution of the stellar component}

Barred galaxies undergo considerable evolution even after the bar and
peanut have formed.  This is due to the fact that bars transfer angular
momentum outwards \cite{LBK}.  Thus the disc/bar component loses
angular momentum to the halo and the bar slows slows down
(e.g. \citeauthor{W}, \citeyear{W}; 
\citeauthor{LC1}, \citeyear{LC1}; \citeauthor{LC2}, \citeyear{LC2};
\citeauthor{HWein}, \citeyear{HWein}; \citeauthor{Abuta}, \citeyear{Abuta}; 
\citeauthor{DS98}, \citeyear{DS98}; \citeauthor{DS00}, \citeyear{DS00};
etc.).  This slow-down is much more important if the halo contributes a
considerable fraction of the mass within a sphere of radius equal to
the disc radius \cite{DS98}.  At the same time the bar becomes
longer and stronger.  A detailed, quantitative description of this
evolution will be given elsewhere.

\subsection{Evolution due to the gaseous component of the disc}

Further types of evolution can be found when the galaxy has a
sizeable gaseous component in its disc.  Thus the spiral driven in the
gas by the bar can evolve, due to the
collisional nature of the gas.  As shown initially by \citeauthor{Sch}
\shortcite{Sch}, this leads to the formation of rings at the main
resonances, outer rings at the outer Lindblad resonance, inner ones at
the iUHR and nuclear ones at the inner Lindblad
resonance (hereafter ILR).  Such rings can be long-lived.

The exchange of energy and angular momentum within the galaxy becomes more
complicated than in cases with no gas, since the gas is now a third
partner in the exchange process.  Thus it
is possible for the bar to speed up somewhat, rather than slow down
(e.g. \citeauthor{FB}, \citeyear{FB}; \citeauthor{BHSF}, \citeyear{BHSF}).

As a response to the bar forcing, the region in and
around the bar is depleted of its gas, which concentrates in two
narrow lanes along the leading 
edges of the bar (e.g. Athanassoula, \citeyear{A92}).  These are in fact
shock loci and, as a result the gas is driven inwards
towards the central region.  If the galaxy has no ILR, then
the gas accumulates in a small region very near the center.  This can
also be the case if the galaxy has an ILR but the sound speed is very
high, of the order of 30 km/sec \cite{EG, PA}.  If, on the
other hand, the galaxy does have an ILR, as
most barred galaxies probably do, and its sound speed is lower than 20
km /sec, as
expected, then the gas occupies a large region
of radius of the order of the ILR radius.  How can the gas be brought 
further inwards to the innermost few parsecs, so that it can
eventually fall on the nucleus and 
feed it and make it an active galactic nucleus? At least two ways have
been so far proposed.  \citeauthor{HS} \shortcite{HS} propose that,
when the central disc or ring becomes gas-dominated, it becomes
gravitationally unstable and breaks into clumps, whose interactions
and collisions may bring the gas further inwards towards the nucleus.  A second
alternative could be bars (or spirals) within bars, as initially proposed by
\citeauthor{SFB} \shortcite{SFB}.  Recent hydrodynamical
simulations, however, question this mechanism, arguing that
secondary stellar bars are unlikely to increase the mass inflow rate
into the galactic nucleus \cite{MTSS}.  Further work on this subject is
necessary to elucidate the properties and the role of the secondary bars.

When the gas reaches the center it will make a strong central
concentration, whose effect may be to destroy the bar, or at least
decrease its strength \cite{FB, BHSF}.  Indeed \citeauthor{HN}
\shortcite{HN} and \citeauthor{HPN} \shortcite{HPN}
have shown that a sufficiently strong central concentration will make
the $x_1$ orbits unstable, so that a large fraction of the phase space
will be occupied by chaotic orbits.  \citeauthor{NSH}
\shortcite{NSH} grew a massive core in the center of an $N$-body
bar and showed that, provided this has a mass larger
than $\sim$5\% of the combined disc and bulge mass, it destroys the
bar.  However, the mass of black holes in disc galaxies \cite{GBBDF,
  FM} is smaller, by an order of magnitude or more, than the mass required by
\citeauthor{NSH}, and thus may not be sufficient to destroy bars.  

\section{Interactions}

Let us now discuss a second engine which can drive evolution in disc
galaxies, namely interactions.  Contrary to bars, which drive a
relatively slow evolution, interactions can drive a fast, and
sometimes violent, evolution.  There is a large variety of types of
interactions 
and of possible results, of which we will review only a few specific
cases.

\subsection{Tails and bridges}
One of the most spectacular results of interactions is the formation
of tails and bridges, such as observed e.g. in the  Antennae (NGC
4038/4039), the Atoms-for-Peace (NGC 7252), or the Mice (NGC
4676).  These structures are formed in a disc galaxy by the tidal
influence of a  
companion galaxy, in particular in direct passages \cite{TooToo}.  In
such cases, the 
angular velocity of the companion is, temporarily, nearly equal to
that of some of the stars in the disc of the target galaxy, and the
effect of this `broad resonance' can make the tidal tails particularly
strong.

\subsection{Ring galaxies}

Another spectacular result of an interaction is the formation of
rings.  A spheroidal companion hitting a target disc galaxy
near-perpendi\-cu\-larly and not far from its center will produce a
density wave in form of an expanding ring \cite{LToo}.  Often a second
ring appears after the first one and sometimes the two are linked by
spokes \cite{AS}.  A nice example of such a structure can be seen in the 
Cartwheel galaxy (A0035-324), which exhibits two rings and a number of spokes
linking them.  Companions of larger mass make primary rings of larger
amplitude and width, which live longer and expand faster than rings
made by less massive companions \cite{APB}.
 
\subsection{Interaction of a barred galaxy and a small spheroidal
  companion} 

I will here summarise some results of a series of simulations of the
interaction between a target barred disc galaxy and its spherical
companion.  The full analysis of these simulations has not yet been
published, although preliminary results can be found in conference
proceedings (\citeauthor{Anob} \citeyear{Anob}, \citeyear{Acamb} and
\citeyear{Atokyo}).  Similar work, but for non-barred disc galaxies, has 
been published by \citeauthor{WMH} \shortcite{WMH}, \citeauthor{HC}
\shortcite{HC} and \citeauthor{VW} \shortcite{VW}.

Let us first consider the case where the companion is initially in a
near-circular orbit on the equatorial plane of the
target and its mass is equal to that of the disc of the target.  
It spirals fast towards the center of the disc, losing only a small
fraction of its mass in the process.  As it approaches the bar, it
perturbs it strongly, so that a number of the particles orbiting in
the bar are pulled towards the companion and the bar is progressively
emptied and destroyed.  During the interaction and subsequent merging
the disc of the target thickens, but also expands, in such a way that it
still remains a disc, albeit somewhat thickened.  
In the final stages of the evolution, after the companion has
reached the center of the target, the bar is totally destroyed and the
disc of the remnant is axisymmetric with a low density region in the
center, which is occupied by the companion.  Thus the companion either
forms a bulge or contributes to a bulge, thus driving evolution of the target
galaxy along the Hubble sequence from later towards earlier type disc galaxies.

Let us now consider a companion of mass equal to one tenth of the mass
of the disc of the target and let its orbit again be near-circular and on the
equatorial plane of the target.  The evolution is totally
different from that described previously for the case of a high mass
satellite.  The 
companion now takes considerably longer to spiral inwards 
and loses a large fraction of its mass in the process.  Some of its
particles stay in the outer parts of the disc forming a spiral
structure; others get trapped in the outer parts of the bar around $x_1$-type
orbits.  

If the companion is initially on a near-circular orbit which forms an
angle with the plane of the disc of the target, then the whole disc of
the target tilts, but is not destroyed, even when the angle is of the
order of 45$^{\circ}$.  For the massive companion case,
the tilt angle is not far from the angle of the orbital plane of the
companion, and it is considerably smaller for the case of the low mass
companion.

\subsection{Mergings and their remnants}
When the encounter between two galaxies is sufficiently
close and slow, then dynamical
friction will bring them nearer and eventually lead to a merging
(e.g. \citeauthor{Too}, \citeyear{Too}).  Following the 
suggestion by \citeauthor{TooToo} 
\shortcite{TooToo} that the merger remnant 
could be an elliptical, many studies have concentrated on
comparing the properties of $N$-body merger remnants with those of
observed elliptical galaxies (\citeauthor{BH}, \citeyear{BH};
\citeauthor{BarnesSaasFee}, \citeyear{BarnesSaasFee}
and references therein).  The results of such comparisons look
encouraging, but many problems have still to be solved.  Thus
the radial projected density profiles display the $r^{1/4}$ form
characteristic of elliptical galaxies over the main body of the
remnant.  In the center-most parts, however, these profiles do not have the
correct form, unless the progenitor discs have little or no gas, as well as a
considerable bulge component \cite{H92, H93, HSH, MH}.  Also the
angular momentum vector does 
not align with the minor axis of the remnant in cases of disc
progenitors of equal mass.  On the other hand, in cases of progenitors with 3:1
mass ratio the misalignment angles are much smaller, consistent with
observations \cite{FIdZ}.  Whether
these shortcomings are due to inadequacies of the present day
modeling procedure, or whether they are pointing to inadequacies of
the simple scenario where ellipticals would be due to merging of two
equal mass galaxies having properties similar to those of disc
galaxies at $z \sim$ 0, is unclear.

Urged by such considerations, two independent studies turned next to multiple
mergings and examined the structure of the resulting
remnants.  Weil \&
Hernquist (\citeyear{WH1}, \citeyear{WH2}) showed that the angular
momentum vector of the
remnant is well aligned with its minor axis, in good agreement with
observations \cite{FIdZ}, while
\citeauthor{AthVo} \shortcite{AthVo}
underlined a number of properties of the $N$-body
remnant which are in agreement with those of elliptical galaxies.  The
two studies are complementary.  The simulations of Weil \& Hernquist,
far superior in particle number, have used only initial conditions
preselected so that all their six galaxies would merge more or less
simultaneously very early on in the simulation.  This could influence
the properties of the remnant, or, more precisely, could give information on
only one type of merger remnant.  On the other hand, Athanassoula \&
Vozikis have a large variety of initial conditions, including both
haloes common to the whole group and individual haloes around each
galaxy, different halo-to-luminous mass ratios and different
kinematics of the galaxies within the group (in virial equilibrium,
collapsing, expanding or rotating).  However, the number of particles
they used was too low to allow them to calculate anything but global
properties of the merger.  Although they checked the robustness of
their results by repeating some of their simulations with double the
number of points, still a larger number of particles would be 
preferable. 

For these reasons I started a series of simulations, as
background jobs of our GRAPE-5 machines \cite{KFMT}, trying to couple
the strong 
points of the two studies, i.e. the large particle number of Weil \&
Hernquist with the larger variety of initial conditions of Athanassoula \&
Vozikis.  Preliminary results from 
this study show two distinct types of remnants.  In most cases the merger
remnant is a spheroidal object, whose properties are similar to those
of the objects analysed by Weil \& Hernquist, i.e. resemble in many
ways elliptical 
galaxies.  There are, however, a few cases in which the disc nature
of one of the progenitors is preserved, so that the remnant
resembles more a Sombrero type S0 galaxy. Further analysis of this
type of remnants, and of the initial conditions that lead to their
formation, is underway.

\end{article}
\end{document}